\documentclass[twocolumn,reprint,superscriptaddress,nofootinbib]{revtex4-1}
\usepackage{graphicx}
\usepackage{dcolumn}
\usepackage{bm}
\usepackage{epsfig}
\usepackage{tabularx}
\usepackage{multirow}
\usepackage{array,multirow}
\usepackage{float}
\usepackage{booktabs}
\usepackage{hyperref}
\hypersetup{colorlinks=true, citecolor=blue, urlcolor=blue, linkcolor=blue}
\begin{document}
\title{Calculation of Coulomb breakup cross sections using a new Coulomb dynamical polarization potential
}
\author{H. M. Maridi}
\email[Corresponding author: 
]{hmaridi@slcj.uw.edu.pl} 
\affiliation{Heavy Ion Laboratory, University of Warsaw, ul.\ Pasteura 5a, 02-093, Warsaw, Poland}
\affiliation{Physics Department, Faculty of Applied Science, Taiz University, Taiz, Yemen}
\author{K. Rusek}
\affiliation{Heavy Ion Laboratory, University of Warsaw, ul.\ Pasteura 5a, 02-093, Warsaw, Poland}
\author{N. Keeley}
\affiliation{National Centre for Nuclear Research, ul.\ Andrzeja So\l tana 7, 05-400 Otwock, Poland}

\begin{abstract}
	A new method for calculating the Coulomb breakup of unstable neutron-rich isotopes at high energies is presented. The calculations employ the eikonal approximation and use a new Coulomb dynamical polarization potential (CDPP), calculated by solving the Schr\"odinger equation for the entire motion of the exotic projectile as a two-body cluster structure using the adiabatic approximation and incorporating excitations to the continuum. 
	Calculations for some exotic isotopes are compared with Coulomb dissociation cross section data and found to be in good agreement.
\end{abstract}

\date{\today}%
\maketitle

\section{\label{sec:int} Introduction}
Recently \cite{Mar21}, we presented a new expression for the Coulomb dynamical polarization potential (CDPP) which was obtained by solving the Schr\"odinger equation for the internal motion of an exotic neutron-rich projectile (considered as a two-body deuteronlike cluster structure) incident on a heavy target nucleus using the adiabatic approximation.
In this work we generalize this CDPP to include both excited states of the core and excitations to the continuum. This generalized CDPP is then used to calculate the differential Coulomb dissociation (CD) cross section as a function of relative excitation energy. These calculations start from the continuity equation using the imaginary part of the CDPP. The eikonal approximation is then used to calculate the CD cross section for a number of weakly-bound neutron-rich exotic nuclei at high incident energies.

\section{\label{sec:theory} Theory}

\subsection{\label{sec:old CDPP} \textit{Continuum based} CDPP}
In Ref.\ \cite{Mar21} we gave the formalism for the scattering of a weakly-bound two-body projectile (p) consisting of a core in its ground state plus a cluster of $n$ valence neutrons from a heavy-ion target and the CDPP was obtained.	
If the core is in an excited state, this expression for the CDPP can be easily generalized by replacing the binding energy of the valence neutron or neutron cluster with respect to the charged core of the projectile, $\varepsilon_0$, with an effective separation energy $\varepsilon^*_0=\varepsilon_0+\varepsilon_{I^{\pi}_c}$, where $\varepsilon_{I^{\pi}_c}$ is the excitation energy of the core state of spin-parity $I^{\pi}_c$.
To solve the Schr\"{o}dinger equations of the system and obtain the CDPP one may use the adiabatic approximation $\Psi ({\bf r,R})\approx \psi ({\bf R}) \phi ({\bf r,R})$, where $\psi ({\bf R})$ refers to the wave function of the center of mass and $\phi ({\bf r,R})$ to that of the relative motion of the projectile; ${\bf R}$ and ${\bf r}$ are the coordinates of the projectile-target and the projectile valence-core systems, respectively. By making the same approximations as in Ref.\ \cite{Mar21} the real and imaginary parts of the CDPP can be given as:
\begin{eqnarray}
\label{eq:CDPP1}
\delta V (R)&=& \varepsilon_0^* \left[ \frac{QG_{0}F_{0}+Q^2 G_{0}F_{0}G_{0}'F_{0}' +Q^2 F_{0}^2 F_{0}'^2}{F_{0}^4 + G_{0}^2 F_{0}^2}-1 \right] \nonumber \\
\delta W (R)&=& \varepsilon_0^* \left[ \frac{Q^2 F_{0}F_{0}' -Q F_{0}^2}{F_{0}^4 + G_{0}^2 F_{0}^2} \right]
\end{eqnarray}
where $F_{0}$ and $G_{0}$ are the regular and irregular Coulomb functions in $\rho = k(R)R$ and $Q(R) = (\mu_p/m_c)(k(R)/\kappa_0)$ with $\kappa_0 = \sqrt{-2\mu_p\varepsilon_0^*/\hbar^2}$ where $\mu_p$ is the core-valence reduced mass, $m_{c}$ the mass of the charged core, and
\begin{equation}
k(R) \approx {\sqrt{\frac{2 m_{c}^2}{\mu_p {\hbar}^2}(V_C ({ R})+\varepsilon_0^*)}}
\end{equation}
is the wave number of the charged core in the field of the target that is associated with the wave function of the internal motion of the projectile, $\phi(r,R)$. It depends parametrically on the Coulomb potential between the projectile and target, $V_C ({ R})$ and is different from the wave number of the center-of-mass motion of the system that describes the motion of the projectile along the Rutherford trajectory, $K=\sqrt{2 \mu (E-\varepsilon_0^*) / \hbar^2 }$ where $E$ is the incident energy of the projectile and $\mu$ is the reduced mass of projectile-target system.
This CDPP (\ref{eq:CDPP1}) depends on the structure of the system but does not depend on the incident energy of the projectile.

Excitations of the projectile to the continuum can be included by adding the continuum energy $\varepsilon$, a continuous variable, to $\varepsilon_0^*$ so that $k$ now becomes a function of $\varepsilon$:
\begin{equation}
\label{eq:kr2}
k(R,\varepsilon) \approx {\sqrt{\frac{2 m_{c}^2}{\mu_p {\hbar}^2}(V_C ({ R})+\varepsilon_0^*+\varepsilon )}},
\end{equation}
making the same approximations as before, and we accordingly obtain the \textit{Continuum based CDPP}, $\delta U (R,\varepsilon)=\delta V (R,\varepsilon)+i\delta W (R,\varepsilon)$:
\begin{eqnarray}
\label{eq:CDPP2}
\delta V (R,\varepsilon)&=& \varepsilon_0^* \left[ \frac{QG_{0}F_{0}+Q^2 G_{0}F_{0}G_{0}'F_{0}' +Q^2 F_{0}^2 F_{0}'^2}{F_{0}^4 + G_{0}^2 F_{0}^2} \right]\nonumber \\ && -\varepsilon_0^* -\varepsilon \nonumber \\
\delta W (R,\varepsilon)&=& \varepsilon_0^* \left[ \frac{Q^2 F_{0}F_{0}' -Q F_{0}^2}{F_{0}^4 + G_{0}^2 F_{0}^2} \right]
\end{eqnarray}
where $Q$ is now also a function of $\varepsilon$ as well as $R$.

We now consider an $E\lambda$-transition ($E1, E2, ...$) from a bound state with angular momentum $l_{0}$ to the $l$-wave continuum, with $l=l_{0}+\lambda$ where ${\lambda}$ is the transition multipolarity.
The initial bound state ${J_0^{\pi}}$ is described by the wave function of the valence neutron in the $l_{0}j_{0}$ orbital relative to the core $I^{\pi}_c$, $\phi^{I^{\pi}_c}_{l_{0}j_{0}}({\bf r})$, whereas the final state (continuum) is represented by the wave function $\phi^{ }_{\varepsilon lj}({\bf r})$, where $\varepsilon$ in the subscript denotes the dependence on the continuum energy. Here we include the notation $I^{\pi}_c$ to indicate that the core is in state $I^{\pi}_c$.
Then, we define a new CDPP that describes this transition as
\begin{equation}
\label{eq:intCDPP}
\delta U^{{E\lambda,I^{\pi}_c}}_{l_{0}j_{0}\rightarrow \varepsilon lj}({\bf R})=\int d\varepsilon \left\langle \phi^{ }_{\varepsilon lj}({\bf r},\varepsilon) \left|  \delta U^{I^{\pi}_c}({\bf R, r},\varepsilon) \right|  \phi^{I^{\pi}_c}_{l_{0}j_{0}}({\bf r}) \right\rangle
\end{equation}
where $\delta U^{I^{\pi}_c}({\bf R, r},\varepsilon)$ can be represented by expansion in Legendre polynomials as
\begin{equation}
\delta U^{I^{\pi}_c}({\bf R, r},\varepsilon)=\sum_{\lambda}\delta U^{I^{\pi}_c}({R},\varepsilon) P_{\lambda}(\cos(\theta_r)).
\end{equation}
Let us define the excitation energy distribution of the $E\lambda$-transition ${l_{0}j_{0}\rightarrow \varepsilon lj}$ as
\begin{equation}
\label{eq:rhoE00}
\rho^{E\lambda,I^{\pi}_c}_{l_{0}j_{0}\rightarrow \varepsilon lj}(\varepsilon)= \left\langle \phi^{ }_{\varepsilon lj}({\bf r},\varepsilon) \left|  P_{\lambda}(\cos(\theta_r)) \right|  \phi^{I^{\pi}_c}_{l_{0}j_{0}}({\bf r}) \right\rangle
\end{equation}
so that we may write
\begin{equation}
\label{eq:intCDPP3}
\delta U^{E\lambda,I^{\pi}_c}_{l_{0}j_{0}\rightarrow \varepsilon lj} ({ R})=\int d\varepsilon \rho^{E\lambda,I^{\pi}_c}_{l_{0}j_{0}\rightarrow \varepsilon lj}(\varepsilon) \delta U^{I^{\pi}_c}({R},\varepsilon).
\end{equation}

The radial wave function of the bound state is given by $\phi^{I^{\pi}_c}_{l_{0}j_{0}}({ r})=u_{l_{0}j_{0}}({ r})/r$, where the single-particle wave function $u_{l_{0}j_{0}}({ r})$ is usually calculated using a single-particle potential model, see for example Ref.\ \cite{Xu94}, by adjusting the Woods-Saxon potential parameters to reproduce the experimental neutron separation energy, taking the excitation energy of the core into account where appropriate. Typical values of the potential radius parameter $r_0$ range from 1.15 to 1.25 fm and the diffuseness $a$ from 0.5 to 0.7 fm. It is, however, important to note that the extracted spectroscopic factor can differ by 20$\%$ or more depending on the choice of parameters \cite{Pra03}. We choose here the potential parameters ($r_0=1.15$ fm, $a=0.5$ fm) suggested by Sauvan \textit{et al.\/} \cite{Sau00} for light neutron-rich nuclei. 
Note that the potential includes a Thomas form spin-orbit term with a strength fixed at 7.0 MeV for all the cases studied here.

The final state (continuum) wave function is usually calculated assuming the plane wave approximation (expanded in spherical Bessel functions and Legendre polynomials) with the radial wave function normalized as $\phi_{\varepsilon lj}(r)=\sqrt{2 \mu_{p} \kappa / \hbar^2 \pi} j_{l}(\kappa r)=u_{\varepsilon lj}({ r})/r$, where $\kappa=\sqrt{2 \mu_{p}\varepsilon}/\hbar$ with relative (continuum) energy $\varepsilon \equiv \varepsilon_{rel}$, and $l$ is the orbital angular momentum of the final state.

Now, by considering the expansions of the bound-state and the continuum plane wave functions,
Eq.\ (\ref{eq:rhoE00}) becomes
\begin{eqnarray}
\label{eq:rhoE01}
\rho^{E\lambda,I^{\pi}_c}_{l_{0}j_{0}\rightarrow \varepsilon lj} (\varepsilon)
&=& \int  d{\bf r} \phi^{}_{\varepsilon lj}({\bf r},\varepsilon)  P_{\lambda}(\cos(\theta_r)) \phi^{I^{\pi}_c}_{l_{0}j_{0}}({\bf r}) \nonumber \\
&=& 4 \pi \sqrt{\frac{2l+1}{2\lambda+1}} \int  d\Omega_r Y_{l_{0}0} (\Omega_r) Y_{\lambda 0}(\Omega_r) Y_{l0}(\Omega_r) \nonumber \\ &\times&
\int_{0}^{\infty} dr u_{\varepsilon lj} \left( r\right) u_{l_{0}j_{0}} \left( r\right)
\nonumber \\
&=& \sqrt{4 \pi(2l_{0}+1)}  \left\langle l_0 0 \lambda 0 | l 0 \right\rangle^2
 \nonumber \\ &\times& \int_{0}^{\infty} dr\ u_{\varepsilon lj} \left( r\right) u_{l_{0}j_{0}} \left( r\right).
\end{eqnarray}
 
Instead of the single-particle wave functions, the asymptotic form for overlap integrals is often used  
\begin{eqnarray}
\label{eq:whit01}
u_{l_{0}j_{0}}({ r})= C_{l_{0}j_{0}} W_{-\eta_0,l_{0}+1/2}(2\kappa_0 r)
\end{eqnarray}
where $W_{-\eta_0,l_{0}+1/2}$ is the Whittaker function and $C_{l_{0}j_{0}}$ is the asymptotic normalization coefficient (ANC). The quantity $\kappa_0=\sqrt{-2 \mu_{p}\varepsilon_0^* }/\hbar$ where $\varepsilon_{0}^*=\varepsilon_0+\varepsilon_{I^{\pi}_c}$ is the effective separation energy of the valence nucleon(s) with respect to the different core states ${I^{\pi}_c}$ of excitation energies $\varepsilon_{I^{\pi}_c}$. When the valence cluster is formed exclusively of
neutron(s) $\eta_0=Z_c Z_v e^2 \mu_p / \hbar^2 \kappa_0 = 0 $ and 
$W_{0,{l_{0}+\scriptscriptstyle\frac{1}{2}}} (2\kappa_0 r)=\frac{2\kappa_0 r}{\pi}k_{l_{0}}(\kappa_0 r)$ 
where $k_{l_{0}}$ is a modified spherical Bessel function and  $h^{(1)}_{l_{0}}$ is a spherical Hankel function of the first kind. 
Then $u_{l_{0}j_{0}}( r)=C_{l_{0}j_{0}}\frac{2\kappa_0 r}{\pi}k_{l_{0}}(\kappa_0 r)$.  
Now by using the Whittaker function for the bound-state wave function, Eq.\ (\ref{eq:rhoE01}) becomes
\begin{widetext}
\begin{eqnarray}
\label{eq:rhoE02}
\rho^{E\lambda,I^{\pi}_c}_{l_{0}j_{0}\rightarrow \varepsilon lj} (\varepsilon)
&=&  \sqrt{4 \pi(2l_{0}+1)} \left\langle l_0 0 \lambda 0 | l 0 \right\rangle^2
\int_{0}^{\infty} dr\ u_{\varepsilon lj} \left( r\right) u_{l_{0}j_{0}} \left( r\right) \nonumber \\
&=&  C_{l_{0}j_{0}}  \sqrt{4 \pi(2l_{0}+1)} \left\langle l_0 0 \lambda 0 | l 0 \right\rangle^2 {2 \kappa_0 \over \pi} \sqrt{2 \mu_{p} \kappa \over \hbar^2 \pi}  \int_0^\infty d{ r} r^2 k_{l_{0}}({\kappa_0 r}) j_{l}({\kappa r})
\nonumber \\ &=& C_{l_{0}j_{0}} \sqrt{4 \pi(2l_{0}+1)} \left\langle l_0 0 \lambda 0 | l 0 \right\rangle^2 \sqrt{2 \mu_{p} \kappa_0 \over \hbar^2 \pi}   
\frac{\kappa^{{\ell}+1/2}}{\kappa_0^{{\ell}+5/2}} \frac{\Gamma(\frac{\ell+\ell_0+3}{2}) \Gamma(\frac{\ell-\ell_0+2}{2})}{\Gamma (\ell+\frac{3}{2})} \nonumber\\ &\times& 
{_{2}F_1} \left( \frac{\ell+\ell_0+3}{2}, \frac{\ell-\ell_0+2}{2},\ell+\frac{3}{2},-\frac{\kappa^2}{\kappa_0^2} \right). 
\end{eqnarray} 
\end{widetext}

In the particular case when the initial bound state has $l_0 = 0$, the radial wave function can be given analytically as the Yukawa form $\phi^{I^{\pi}_c}_{l_{0}j_{0}}({ r})=\frac{u_{l_{0}j_{0}}({r})}{r}=C_{s_{1/2}} \frac{e^{-\kappa_0 r}}{r}$, so that for the $E1$ transition to the $p$-wave continuum,  we obtain the following analytical function:

\begin{equation}
\label{eq:rhoE0sp}
\rho^{{E1},I^{\pi}_c}_{s\rightarrow p}(\varepsilon)
= C_{s_{1/2}} \sqrt{\frac{8 \mu_{p} \kappa}{\hbar^2}}\frac{1}{\kappa^2}\left( \frac{-\kappa_0 \kappa}{\kappa_0^2+\kappa^2} +\tan^{-1}\left( \frac{\kappa}{\kappa_0} \right) \right)
\end{equation}

which will be especially useful since the valence neutron(s) are in a relative $s$-state in the ground state of many light exotic projectiles of interest.

\subsection{\label{sec:CD} Coulomb dissociation using the CDPP}
This new CDPP (\ref{eq:intCDPP3}) has many potential applications to the scattering and reactions of weakly-bound neutron-rich nuclei. Here we use it to calculate Coulomb dissociation, keeping other applications for future publications.
Coulomb breakup or dissociation can take place when a high energy (several hundred MeV/nucleon) projectile impinges on a heavy target and is excited by absorbing virtual photons from the time-dependent Coulomb field. Under these circumstances the electromagnetic excitation is dominated by dipole excitation \cite{Pra03}. The corresponding differential cross section for dipole excitation decomposes into an incoherent sum of components corresponding to the different core states populated by neutron removal. For each core state the cross section further decomposes into an incoherent sum over contributions from the different allowed angular momenta of the valence neutron in its initial state. In these calculations the projectile is assumed to have a core plus valence neutron structure. Choosing the lowest core states there are various ways to couple the spins of the core and the valence neutron to the total angular momentum and parity of the projectile \cite{Hei17}.
Note that for heavy targets the core is assumed to act as a spectator so that the projectile core remains in its initial state after removal of a neutron \cite{Han03}.

We start from the usual formula for the absorption cross section derived from the continuity equation
\begin{equation}
\label{eq:partsigma0}
\sigma^{ }_\mathrm{abs}  =- \frac{2}{\hbar v} \left\langle  \psi^{(+)}_{K} ({\bf R}) \left|  W ({ R}) \right|  \psi^{(+)}_{K} ({\bf R}) \right\rangle
\end{equation}
where  $W({ R})$ is the imaginary potential, $\psi^{(+)}_{K}$ the usual distorted wave function, $v$ the relative velocity, and $K$ the wavenumber in the center of mass system.
By considering the imaginary CDPP (\ref{eq:intCDPP}) that describes the ${E\lambda}$-transitions we can write the total absorption due to Coulomb dissociation and excitation by taking a summation over the ${E\lambda}$-transitions, the different core states $I^{\pi}_c$,  and contributions from different angular momenta $l_{0}j_{0}$ of the valence neutron in its initial state for each core state:
\begin{widetext}
\begin{eqnarray}
\label{eq:partsigma}
\sigma^{ }_\mathrm{CD}  &=&- \frac{2}{\hbar v} \left\langle  \psi^{(+)}_{K} ({\bf R}) \left|  \delta W^{ }_{ }({ R}) \right|  \psi^{(+)}_{K} ({\bf R}) \right\rangle
\nonumber \\ &=&
- \frac{2}{\hbar v} \int d\varepsilon \sum_{I^{\pi}_c} \sum_{\lambda} \sum_{l_{0}j_{0}} \sum_{lj} \rho^{E\lambda,I^{\pi}_c}_{l_{0}j_{0}\rightarrow \varepsilon lj} (\varepsilon) \left\langle  \psi^{(+)}_{K} ({\bf R}) \left|  \delta W^{I^{\pi}_c}({ R},\varepsilon) \right|  \psi^{(+)}_{K} ({\bf R}) \right\rangle .
\end{eqnarray}
\end{widetext}
Then, the differential cross section for the $E\lambda$-transition $l_{0}j_{0}\rightarrow \varepsilon lj$ with the core in state $I^{\pi}_c$ is
\begin{widetext}
\begin{equation}
\label{eq:partsigma1}
\frac{d\sigma^{I^{\pi}_c}}{d\varepsilon} (E\lambda,{l_{0}j_{0}\rightarrow \varepsilon lj}) = - \frac{2}{\hbar v}  \rho^{E\lambda,I^{\pi}_c}_{l_{0}j_{0}\rightarrow \varepsilon lj} (\varepsilon) \left\langle  \psi^{(+)}_{K} ({\bf R}) \left|  \delta W^{I^{\pi}_c}_{ }({ R},\varepsilon) \right| \psi^{(+)}_{K} ({\bf R}) \right\rangle ,
\end{equation}
\end{widetext}
which is similar to the well-known Hussein--McVoy formula and the similar formulae for the inclusive breakup \cite{Hus85,Ram21}.
Introducing
\begin{eqnarray}
\label{eq:unnormsigma}
\hat{\sigma}^{I^{\pi}_c}_{ }(\varepsilon)&=&- \frac{2}{\hbar v} \left\langle \psi^{(+)}_{K} ({\bf R}) \left|  \delta W^{I^{\pi}_c}_{ }({\bf R},\varepsilon) \right|  \psi^{(+)}_{K}({\bf R}) \right\rangle
\end{eqnarray}
as the total absorption cross section for the $I^{\pi}_c$ core state,
the breakup cross section for the core in state $I^{\pi}_c$ and the $E\lambda$-transition from ${l_{0}j_{0}}$ is
\begin{equation}
\label{eq:con11}
\sigma^{I^{\pi}_c}_{E\lambda,l_{0}j_{0}}
= \sum_{lj}\int_0^{\infty} d\varepsilon \rho^{E\lambda,I^{\pi}_c}_{l_{0}j_{0}\rightarrow \varepsilon lj} (\varepsilon) \hat{\sigma}^{I^{\pi}_c}_{ }(\varepsilon).
\end{equation}

Equations (\ref{eq:partsigma1}) and (\ref{eq:con11}) can be generalized to account for excitations of the target. An imaginary nuclear potential can be added to account for any other reaction channels. From Equation (\ref{eq:con11}) and using the virtual photon method \cite{Ber88} one may obtain the photo-absorption cross section $\sigma_{\gamma n}$, the radiative neutron capture cross section $\sigma_{n \gamma}$, the dipole response function $dB/d\varepsilon$, and the double differential cross section $d^2 \sigma / d \Omega d\varepsilon$.
These calculations can also be generalized to consider transitions between two bound states by replacing the continuum wave function
$u_{\varepsilon lj} (r)$ with a bound state wave function $u_{lj} (r)$ similar to $u_{l_{0}j_{0}} (r)$.

Equation (\ref{eq:unnormsigma}) can be solved by the method of partial wave expansion and the complete scattering amplitude is obtained by summing over all the partial wave scattering amplitudes.
At high energies, i.e.\ hundreds of MeV/nucleon, the wave function will oscillate rapidly and the calculation of scattering wave functions for each partial wave becomes more complicated. In the eikonal approximation and the optical limit of the Glauber theory \cite{Gla59}, the wave function of the projectile-target system can be written in terms of the total potential, $V+iW$, as
\begin{equation}
\label{eq:weik11}
\psi_K ({\bf R}) = \exp\left( iKz+\frac{1}{i \hbar v}\int_{-\infty}^{z} (V(\mathbf{b},{z}')+iW(\mathbf{b},{z}')) d{z}'\right)
\end{equation}
where ${\bf R}=(x,y,z)=(\mathbf{b},z)$ and ${\bf b}$ is the impact parameter.
Equation (\ref{eq:unnormsigma}) may then easily be written as
\begin{eqnarray}
\label{eq:j44}
\hat{\sigma}^{I^{\pi}_c}_{ }(\varepsilon)
 &=& \int d\mathbf{b}  \left[1 -e^{\frac{2}{\hbar v}\int_{-\infty}^{\infty}  \delta W^{I^{\pi}_c}_{ }({\bf b},\acute{z},\varepsilon) d\acute{z}} \right]
\end{eqnarray}
where $ d\mathbf{b}=2 \pi b db$.
At this point we should include the correction of the impact parameter due to Coulomb deflection of the particle trajectory: $b'=a_0 + \sqrt{a_0^2 +b^2}$, where $a_0=Z_p Z_t e^2 / mv^2$ is half the distance of closest approach in a head-on collision of point charged particles. The relativistic correction to the kinematics can be appropriately made if one replaces the quantity $a_0$ with $a_0=Z_p Z_t e^2 / \gamma mv^2$,  where $\gamma=1/\sqrt{1-v^2/c^2}$ is the Lorentz factor \cite{Ber03}.
Since the phase in Eq.\ (\ref{eq:j44}) is integrated over $z$ it is invariant under Lorentz transformations, so no further relativistic corrections to the dynamical equations are required \cite{Mos19}.

\section{\label{sec:app} Application to some reactions}
We now apply our model to calculate the Coulomb dissociation cross sections for some exotic projectiles ($^{11}$Be, $^{15}$C, $^{17}$C) incident on a lead target at a few hundreds of MeV/nucleon. Our results using both the single particle (s.p.)\ wave function $u_{l_{0}j_{0}}$ and the analytical formula (\ref{eq:rhoE02}) which assumes a Whittaker wave function for the bound states are summarized in Table \ref{tab:cd}. 
All of these calculations consider only the $E1$ transition to the continuum.
In addition, we define $R_{\sigma}= {d\sigma_\mathrm{exp} \over d\varepsilon}/{d\sigma_\mathrm{th} \over d\varepsilon}$ which is the ratio of the experimental to the theoretical differential cross section using the best fit in the energy range of the data. This ratio can be compared with the theoretical and empirical spectroscopic factors listed in the last two columns of Table \ref{tab:cd}. 
The experimental data are known to be affected by the experimental resolution, so, using the method of Ref.\ \cite{Mor20}, the calculated cross sections were convoluted with the detector response function obtained from simulated spectra. Since the Coulomb dissociation data of $^{11}$Be, $^{15}$C, and $^{17}$C were all measured at GSI and their experiments employed a similar setup \cite{Pal03,Pra03} we use the energy resolution obtained in Ref.\ \cite{Pal03} for all calculations.	
The cross section is calculated by integration up to a continuum energy of 10 MeV for most cases. 
\begin{table*}
	\centering
	\caption{\label{tab:cd} Calculated Coulomb dissociation cross sections for several exotic projectiles (with a core-neutron structure) incident on lead targets for different core excitation states using single-particle (s.p.) and Whittaker functions for the \ bound-state wave functions $u_{l_{0}j_{0}}$. 
The ratios $R_{\sigma}$ (see text) are compared with the spectroscopic factors extracted from experiment and from Shell-model calculations.}
	\begin{tabularx}{\linewidth}{@{\extracolsep{4pt}}cccccccccccc>{\centering\arraybackslash}X@{}}
		\hline\noalign{\smallskip}
		Proj.&E/n &$J^{\pi}$ &Core&$I^{\pi}_{c}$ &${l_{0}j_{0}}$ &$\sigma_\mathrm{exp}$ (mb)&\multicolumn{2}{c}{$\sigma_{th}$ (mb)}&\multicolumn{2}{c@{}}{$R_{\sigma}$}&\multicolumn{2}{c@{}}{Spectroscopic factors} \\
		\cline{8-9} \cline{10-11} \cline{12-13} \noalign{\smallskip}
		&  &  & &  &  &  &s.p.&Whitt. &s.p. &Whitt.&Shell model &experiment \\
		\hline\noalign{\smallskip}
		$^{11}$Be&520 &$1/2^{+}$&$^{10}$Be &$0^{+}$ &$s_{1/2}$ &605$\pm$30 \cite{Pal03}   &1563   &1649   &0.40 &0.40 &0.74 \cite{Aum00} & 0.36, 0.60 \cite{Tim99}; 0.61(5) \cite{Pal03}; 0.72(4) \cite{Fuk04}; 1.0(2) \cite{Nak94}  \\
		&  & &  &$2^{+}$ &$d_{5/2}$ &  &34.4  &58.2 &  & & & \\
		\hline\noalign{\smallskip}
		$^{15}$C&605  &$1/2^{+}$ &$^{14}$C  &$0^{+}$ &$s_{1/2}$     &324$\pm$15 \cite{Pra03}  &570 &643 &0.52 &0.48 &0.98 \cite{Ter04}, 0.83 \cite{Sau00}  &0.91(6), 0.72(5)  \cite{Nak09}; 0.73(5), 0.97(8) \cite{Pra03} \\
		&  & &  &$1^{-}$ &$p_{1/2}$ &36$\pm$3 \cite{Pra03} &8.2   &17.1  & &  & 1.03 \cite{Sau00}&1.3(1)  \cite{Pra03} \\
		&  & &  &  &$p_{3/2}$ &  &8.3   &17.9 & & &  0.16 \cite{Sau00}& \\
		&  & &  &$0^{-}$ &$p_{1/2}$  &  &4.8 &10.5 & & &0.46 \cite{Sau00,Ter04} & \\
		&  & &   &$2^{-}$ &$p_{3/2}$     &  &3.9 &8.8 & & &0.016 \cite{Ter04} & \\
		\hline\noalign{\smallskip}
		$^{17}$C&496  &$3/2^{+}$ &$^{16}$C  &$0^{+}$ &$d_{3/2}$     &$9^{+15}_{-9}$ \cite{Pra03}&607 &854 &0.015 &0.011 &0.035 \cite{Sau00}& \\
		&  & &    &$2^{+}$ &$s_{1/2}$     &62$\pm$7 \cite{Pra03}  &118.9   &145.5  &0.28 &0.23 &0.16\cite{Sau00,Mad01}  &0.23(8),26(14) \cite{Pra03}\\
		&  & &     &   &$d_{5/2}$     &  &96.8    &162.1 &0.23  &0.14 &1.41 \cite{Sau00}, 1.44 \cite{Mad01}& 0.6(4),1.6(6) \cite{Pra03} \\
		&  & &    &$0^{+}$ &$d_{3/2}$     &25$\pm$7 \cite{Pra03} &33.9 &60.5 & & & & \\
		&  & &    & $4^{+}$ &$d_{5/2}$     & &15.5 &30.1 & & &0.76 \cite{Sau04,Mad01}& \\
		\hline\noalign{\smallskip}	  
	\end{tabularx}
\end{table*}
\begin{figure}[tb]
	\centering
	\includegraphics[width=0.45\textwidth,clip=]{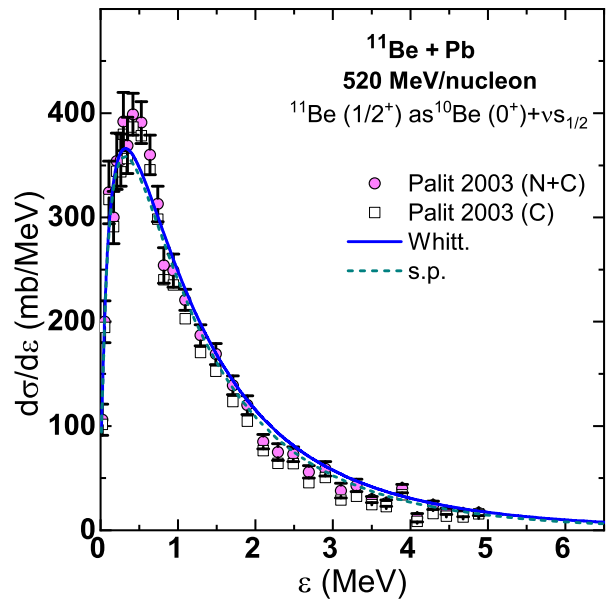}
	\caption{\label{fig:11Be} Coulomb dissociation cross section for a $^{11}$Be projectile incident on a lead target at 520 MeV/nucleon. 
	The solid symbols represent the data of Ref.\ \cite{Pal03} and the open symbols the data after removal of the nuclear contribution. 
	The solid and short-dashed lines represent calculations using Whittaker and single-particle wave functions for the bound state wave function	$u_{l_{0}j_{0}}$, respectively. The calculations have been convoluted with the experimental resolution. See text for details.}
\end{figure}

We take as our first example the Coulomb dissociation of $^{11}$Be. The ground state of $^{11}$Be is $1/2^{+}$, so the $0^{+}$ ground state of the $^{10}$Be core is coupled to an $s$-wave valence neutron and the $2^{+}$ first excited state of the core (3.368 MeV) is coupled to a $d$-wave valence neutron. The spectroscopic factors for the $\left| ^{10}\mathrm{Be}(0^+)\otimes \nu 2s_{1/2}\right\rangle $ single-particle configuration obtained from shell model calculations and deuteron stripping and pick-up reactions range from about $0.4$--$0.8$, see for example \cite{Aum00} and references therein. 
An analysis of the Coulomb dissociation of a $^{11}$Be projectile with an energy of 520 MeV/nucleon impinging on lead and carbon targets gives a spectroscopic factor for the $\left| ^{10}\mathrm{Be}(0^+)\otimes \nu 2s_{1/2}\right\rangle $ single-particle configuration of 0.61(5) \cite{Pal03}. 
These data \cite{Pal03} include both Coulomb and nuclear contributions and the Coulomb breakup spectrum
can be extracted by subtracting the nuclear contribution estimated using the data taken with the carbon target in the following way
\begin{equation}
\label{eq:scale}
\frac{d\sigma_{CD}}{d\varepsilon}=\frac{d\sigma}{d\varepsilon}(\mathrm{Pb})-\Gamma\frac{d\sigma}{d\varepsilon}(\mathrm{C})
\end{equation}
where $\Gamma$ is a scaling factor. 
Assuming the peripheral nature of the nuclear excitation, $\Gamma$ can be taken as the ratio of the sum of the radii of the target and the projectile which is 1.8 \cite{Fuk04,Nak94}.
Other estimates of $\Gamma$ for this reaction may be found in Refs.\ \cite{Fuk04,Nak94,Pal03}.
The effect of the nuclear contribution on this reaction has also recently been studied in Ref.\ \cite{Mor20} by comparing the Equivalent Photon Method (EPM) and continuum discretized coupled channels (CDCC) calculations. 

The analysis of the data using the present formalism, with the Yukawa and s.p. wave functions, gives us calculated differential cross sections in good agreement with the data, as shown in Fig.\ \ref{fig:11Be}, and a cross section ratio of about 0.4 is obtained, which is similar to the spectroscopic factor of Ref.\ \cite{Tim99}.  
It is clear that the contribution from the $\left| ^{10}\mathrm{Be}(2^+)\otimes \nu 1d_{5/2}\right\rangle $ configuration is small. We expect that the contribution from the higher $1^{-}$ and $2^{-}$ core states with $p$-wave neutrons will also be small.
\begin{figure}[tb]
	\centering
	\includegraphics[width=0.45\textwidth,clip=]{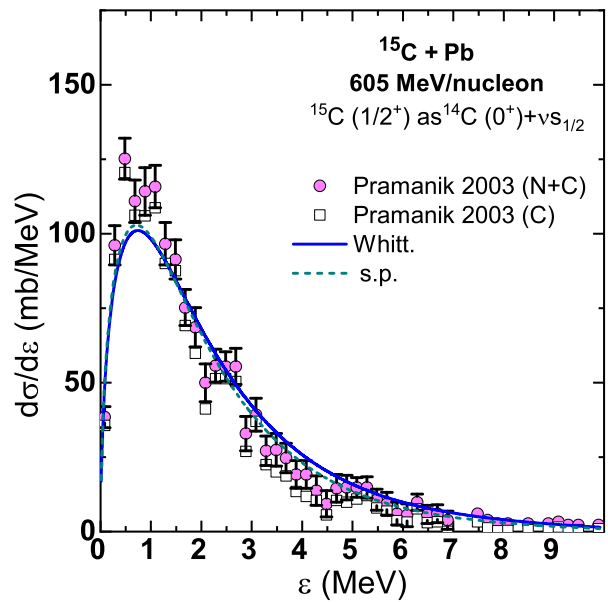}
	\caption{\label{fig:15C} The same as Fig. \ref{fig:11Be} but for a 605 MeV/nucleon $^{15}$C projectile. The data are from Ref.\ \cite{Pra03}. The calculations have been convoluted with the experimental resolution.}
\end{figure}

The one-neutron separation energy of $^{15}$C is 1.218 MeV and its ground-state spin-parity is $1/2^+$, so we suppose a $2s_{1/2}$ neutron coupled to a $^{14}$C($0^+$) core.
Figure \ref{fig:15C} presents calculations of the Coulomb breakup of $^{15}$C incident on a Pb target at 605 MeV/nucleon \cite{Pra03}. The peak of the relative energy spectrum occurs at a value of 0.7 MeV.
The original data \cite{Pra03} that include both Coulomb and nuclear contributions as well as the data obtained by subtracting the nuclear contribution are plotted by the filled dots and open squares, respectively. 
The calculated total Coulomb dissociation cross section for $^{15}$C$\rightarrow$$^{14}$C+$n$+$\gamma$ is about 570 mb using the s.p.\ wave function and 643 mb using the Yukawa wave function for the $|^{14}\mathrm{C}(0^+)\otimes \nu 2s_{1/2}>$ configuration, and the corresponding ratios of the experimental to calculated cross sections are about 0.52 and 0.48, respectively. 
These ratios are compared with the spectroscopic factors calculated by the shell model \cite{Ter04} or extracted from Coulomb dissociation experiments \cite{Nak09,Pra03} in Table \ref{tab:cd}.
There are many excited states in $^{14}$C at energies between 6--7 MeV and previous calculations with the core in these states, see Ref.\ \cite{Pra03}, gave small cross sections compared to the observed value of 36(3) mb. For the 6.09 MeV ($1^-$), 6.90 MeV ($0^-$), and 7.34 MeV ($2^-$) core excited states the possible neutron orbitals are $p_{1/2}$ or $p_{3/2}$.
Our calculation using the s.p.\ wave functions gives 4.2, 1.1, and 1 mb for dipole transitions to the $d$-continuum and 12.4, 3.7, and 3 mb for transitions to the $s$-continuum for the $1^-$, $0^-$, and $2^-$ core states, respectively. 
The calculation using the Whittaker wave functions gives 4.7, 1.3, and 1.1 mb for dipole transitions to the $d$-continuum and 30.2, 9.2, and 7.8 mb for transitions to the $s$-continuum for the $1^-$, $0^-$, and $2^-$ core states, respectively.
If we multiply these cross sections by the corresponding spectroscopic factors from shell model calculations \cite{Sau00,Ter04} this gives us values of 12.1 mb and 25.4 mb from the calculations using s.p. and Whittaker functions, respectively, whereas the experimental value is 36(3) mb \cite{Pra03}. The remaining cross section may correspond to the $^{14}$C core in the other three states at 6--7 MeV.
\begin{figure}[tb]
	\centering
	\includegraphics[width=0.45\textwidth,clip=]{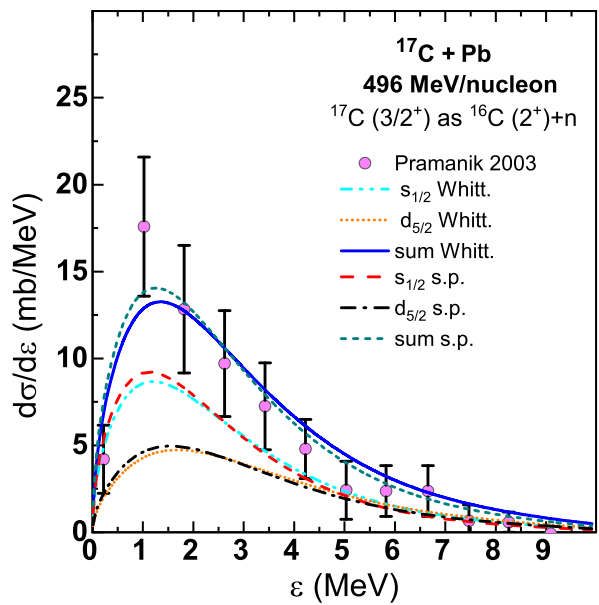}
	\caption{\label{fig:17C} Coulomb dissociation cross sections for a 496 MeV/nucleon $^{17}$C projectile incident on a lead target. The data are from Ref.\ \cite{Pra03}. The calculations have been convoluted with the experimental resolution.}
\end{figure}

Calculations for different core excitations were performed for the $^{17}$C Coulomb breakup since this yields the $^{16}$C core mainly in excited states. The differential Coulomb dissociation cross section for $^{17}\mathrm{C} \rightarrow \protect{^{16}\mathrm{C}}(2^{+})+n$ is reported in Ref.\ \cite{Pra03}.
As shown in Fig.\ \ref{fig:17C}, the data are well reproduced by our calculation using a Yukawa wave function for the $|^{16}\mathrm{C}(2^+)\otimes \nu 2s_{1/2}>$ configuration with an integrated cross section of 60 mb, similar to the reported one of 62(7) mb. The peak position is also well reproduced.
The calculations using the s.p.\ wave functions and Whittaker wave function can fit the data with cross section ratios of about 0.28 and 0.23 for the $s_{1/2}$ and  0.23 and 0.14 for $d_{5/2}$-orbits, respectively. The contribution of both orbits to the cross section of the $^{16}$C($2^{+}$) excited state is not known experimentally, only the total cross section of 62(7) mb. There are thus many combinations which can fit the data, and the cross section ratio for $s_{1/2}$ or $d_{5/2}$ can be varied from zero to 0.5--0.6.
As reported in Ref.\ \cite{Pra03}, the measured cross sections show that 64(9) \% of the cross section corresponds to the ($2^{+}$, 1.766 MeV) core state,  27(9) \% to higher core excited states at 3--4 MeV and a small part of the cross section leaves the core in its $0^{+}$ ground state \cite{Pra03}.
As shown in Table \ref{tab:cd}, our calculation using the s.p.\ wave functions for the core in its ground state with a $1d_{3/2}$ valence neutron gives a cross section of 153 mb for the transition to the $f$-wave and 454 mb for the transition to the $p$-wave continuum to give a total of 607 mb, and the resulting ratio is very small (about 0.015), close to the value expected from a shell model calculation \cite{Mad01}. 
Calculations using the Whittaker wave functions give a cross section of 854 mb with ratio of 0.011.
For the higher states, the spectroscopic factor is unknown for the $1d_{3/2}$ valence neutron plus the ($0^{+}$, 3.027 MeV) core state and is 0.76 for the $d_{5/2}$ valence neutron in the ($4^{+}$, 4.142 MeV) core state; our calculations using s.p. and Whittaker functions give about 15 mb and 30 mb for the cross section of $4^{+}$ state, respectively, whereas the measured value is 25(7) mb \cite{Pra03}. 

\section{\label{sec:summary} Summary and conclusions}
In summary, we have presented a new method to calculate the Coulomb dissociation of exotic nuclei using an extended version of a recent model of the CDPP taking into account excited states of the core and excitation to the continuum. Breakup cross sections for two-body exotic projectiles may be calculated using this new version of the CDPP which depends on the relative excitation energy of the continuum.
The method was used to calculate the differential and integrated Coulomb breakup cross sections for several exotic neutron-rich nuclei incident on lead targets at a few hundreds of MeV/nucleon.
The calculations at these high energies were performed using the eikonal approximation and the results are in good agreement with the data.
The calculations could easily be generalized and applied to calculate neutron removal and gamma capture cross sections.
The new, extended CDPP may also be applied to low energy scattering and reaction data via the optical model or distorted wave Born approximation or coupled channel formalisms and has many potential applications.

\bigskip
\section*{Acknowledgments}
We thank Professor Antonio Moro for his help in calculating the energy convolution of the calculated cross sections.
This work was funded by the Polish National Agency for Academic Exchange (NAWA) within the Ulam Programme under grant agreement No. PPN/ULM/2019/1/00189/U/00001.

\end{document}